\documentclass[12pt,twoside]{article}
\usepackage{fleqn,espcrc1}

\usepackage{graphicx}
\usepackage[figuresright]{rotating}

\newcommand{\AmS}{{\protect\the\textfont2
  A\kern-.1667em\lower.5ex\hbox{M}\kern-.125emS}}

\begin{document}

\begin{center}
\vspace{4 mm}
REVIEW OF INDIRECT METHODS USED TO DETERMINE THE $^1S_0$ 
       NEUTRON-NEUTRON SCATTERING LENGTH \\
\vspace{4 mm}

C.R. HOWELL\footnote{This work was supported in part by the U.S. Department of 
		Energy, Office of High Energy and Nuclear Physics, 
                under Grant No. DE-FG02-97-ER41033} \\
Duke University and Triangle Universities Nuclear Laboratory \\
Durham, NC  USA \\
\vspace{5 mm}
\end{center}


\begin{abstract}
Calculations with realistic potential models show that a 1\% change in 
the $^1S_0$ nucleon-nucleon (NN) potential
strength changes the scattering length by 30\%.  It is this high sensitivity
that makes the  $^1S_0$ NN scattering lengths important
in quantifying the amount to which charge
symmetry is broken in the strong nuclear force.  
Of the three $^1S_0$ NN scattering lengths, the value of the
neutron-neutron scattering length $a_{nn}$ is the most uncertain.
A number of reactions that produce two neutrons with low relative energy have
been used to determine $a_{nn}$. However, the two reactions that give $a_{nn}$ 
with the least theoretical uncertainty are pion-deuteron capture 
($\pi^-+d \rightarrow n+n+\gamma$) and neutron-deuteron breakup 
($n+d \rightarrow n+n+p$).  Curiously, the values obtained using these 
two reactions are significantly different.  In this talk the experimental 
techniques and theory used to determine $a_{nn}$ for each of these 
popular reactions will be reviewed.  In addition, the results of the two 
most recent {\em nd} breakup experiments will be reported.
\end{abstract}

\section{INTRODUCTION}

Most observed charge-symmetry-breaking (CSB) effects can be 
explained as being due to the differences in the
masses (QMD) and electric charges of the $d$ and $u$ quarks \cite{Mil90}.
These differences are manifested by the mass splitting in
hadronic isospin multiplets and in the values of the $\rho$-$\omega$ 
and $\pi$-$\eta$
mixing amplitudes within meson-exchange potentials \cite{Mil90}. 

The QMD leads to a difference in the masses of the neutron and proton and 
many other hadrons, and 
to a difference between the neutron-neutron ($nn$) and
proton-proton ($pp$) $^1S_0$ scattering lengths, 
$\Delta a = a_{nn}$-$a_{pp}$ \cite{Mil90}.  Therefore, an experimental 
determination of the scattering length difference, $\Delta a$, gives a direct
measure of CSB and can be related to the QMD.
This high sensitivity of the scattering lengths to details of the nuclear 
force at the quark level also is
reflected in nucleon-nucleon (NN) potential 
models that are based on meson-exchange
phenomenology.
For example, for most realistic NN 
potential models, a 1\% change in the potential strength results in a 30\% 
shift in the value of the calculated scattering length. This enormous
sensitivity of the NN scattering length to the potential strength can
be understood to first order using effective range phenomenology.    
For a square well potential, the fractional change in the scattering
length due to a small change in the potential depth is given by

\begin{equation}
\frac{\delta a_{nn}}{a_{nn}} = 1.23 \left( \frac{a_{nn}}{r_{nn}} \right) \frac{\delta V}{V}.
\end{equation}

Using typical values of $a_{nn} = 18.8 fm$ and $r_{nn} = 2.8 fm$, implies
that a 1\% change
in the depth of the square well potential will result in about a 10\% shift in
the value of $a_{nn}$.

The $a_{pp}$ has
been measured directly to high precision (of order 0.01 fm) using 
two-nucleon scattering.  However, there is a relatively large uncertainty
of about $\pm 0.4 fm$ in correcting the measured value of $a_{pp}$ for 
electromagnetic ({\em em}) effects, which are sizeable for this parameter.  
Consequently, the main uncertainty
in determining the nuclear part of $a_{pp}$ is due to the theoretical 
uncertainties in the factors 
used to relate the measured value of $a_{pp}$, which includes all {\em em}
effects, to the purely nuclear part of the scattering length.
The situation is quite
different for $a_{nn}$.  For technical reasons direct measurements of 
$a_{nn}$ using free
neutrons have never been successfully executed.  Up to now, all
determinations  for $a_{nn}$ have been based on studies of 
reactions with at least three particles in the exit channel 
\cite{Sla89,How98,Gon99,Huh00}. In this paper, the results from measurements
of neutron-deuteron ({\em nd}) breakup and pion-deuteron ($\pi^-d$)
capture are reviewed. 
Studies using these reactions were chosen for review because 
historically they have provided the most trusted determinations
of $a_{nn}$.  After the review of two recent kinematically complete  
{\em nd} breakup measurements, some concluding remarks are made and a list of 
recommended next steps presented.

\section{SUMMARY OF $a_{nn}$ RESULTS AND METHODS }

\begin{figure}[htb]
\includegraphics*[width=150mm]{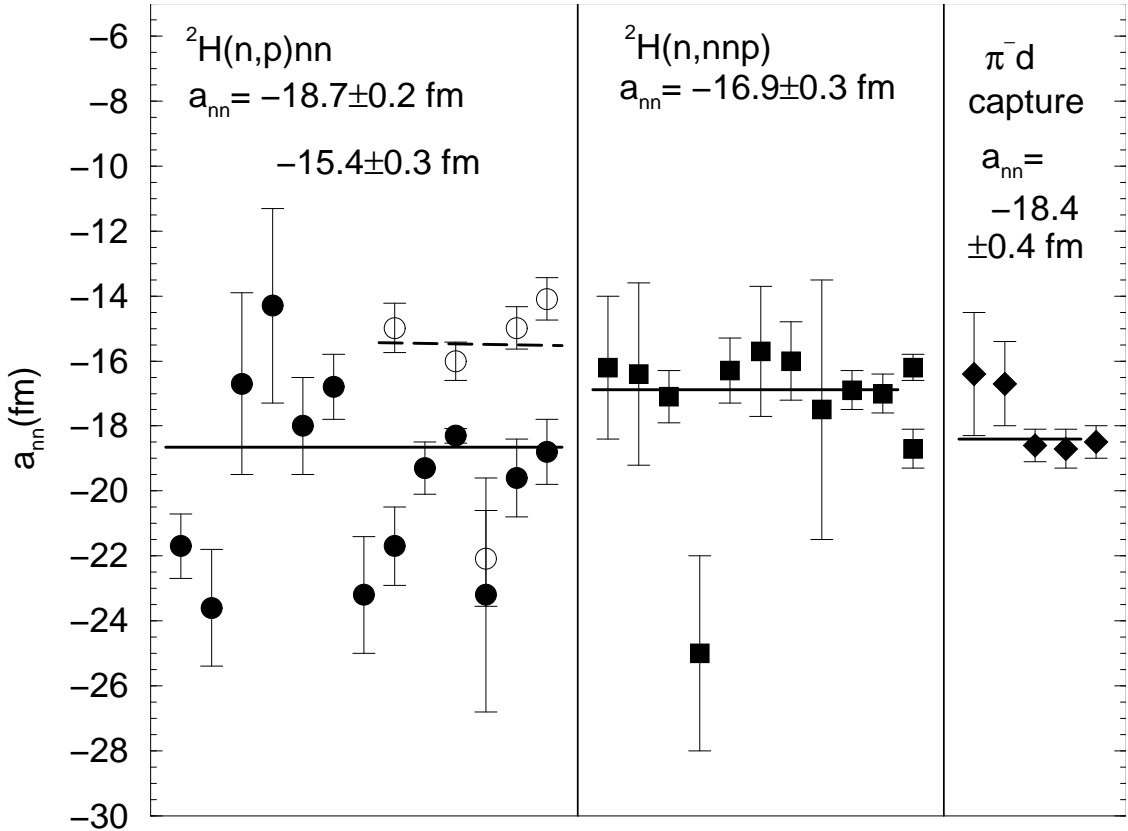}
\caption{Summary of indirect determinations of $a_{nn}$ from experiments
     reported between 1964 and the present.  The data points for each
     reaction type are plotted on the horizontal axis in nearly chronological
     order and are as listed in Tables~\protect\ref{tab:ndbu} and
     \protect\ref{tab:pi_d}. Five of the experiments reanalyzed by
     Tornow {\em et al.} \protect\cite{Tor96} are shown as open circles.
     The average values given at the tops of the sections for kinematically
     complete {\em nd} breakup measurements and the $\pi^-d$ capture
     experiments don't include the results reported since 1997. The second
     average value in the left section is for the values obtained in
     the reanalysis \protect\cite{Tor96}.}
 \label{fig:ann_summary}
\end{figure}

In this section the results for $a_{nn}$ are summarized for studies
done using {\em nd} breakup and $\pi^-d$ capture measurements.
This review will cover only experiments
reported between 1964 and the present.
In all cases $a_{nn}$ was determined by fitting the measured cross section
for the {\em nn} final-state interaction (FSI), which has a maximum
value when the relative momentum of the two interacting neutrons is
nearly zero.  In the data analysis of these experiments, the relative
energy between the two neutrons was typically between 0 and 500~keV.
This wide integration range was required to obtain sufficient statistical
accuracy in the $a_{nn}$ determination.
A graphical summary of the 
situation is shown in Fig.~\ref{fig:ann_summary}. The graph is divided
into three sections.  The left most section contains the results from
cross-section measurements of kinematically incomplete {\em nd} breakup 
experiments. Five of the experiments reanalyzed by Tornow {\em et al.}
\cite{Tor96} using modern theory are displayed as open circles.  
The middle section contains results from kinematically
complete {\em nd} breakup experiments. The results from $\pi^-d$ capture
measurements are in the right section.  The solid line represents the 
stastically weighted average of the data points for each reaction type from
1964 through 1997.
The dashed line in the left section is the average of the reanalyzed data. 

\noindent
Some details of the experiments from which the data in 
Fig.~\ref{fig:ann_summary} are taken are given in the tables in this
section.
Each experiment is categorized according to the number of kinematic quantities measured.
The two broad types of experiments are kinematically complete (KC) and 
kinematically incomplete (KI).
The experiments are tagged according to the number of measured kinematic parameters, 
because the level of  
the kinematic contraints imposed in the experiment can significantly
impact the signal-to-background in the measurement and the theory
used in the data analysis.

\begin{table}[htb]
\caption{Survey of {\em nd} breakup cross-section measurements that were 
   used to determine $a_{nn}$.  The survey covers the period between 1964
   and the present. The main features of the analyses are given in Table~\ref{tab:details}.}
\label{tab:ndbu}
\newcommand{\cc}[1]{\multicolumn{1}{c}{#1}}
\begin{tabular}{|llllll|}
\hline
      &       & \cc{$E_{lab}$} & \cc{Measured}   & \cc{Analysis}  & $a_{nn} \pm \Delta a_{ann}$ \\
ref.  & year  & \cc{(MeV)}     & \cc{Kinematics} & \cc{Details}   & ~~~~(fm)                   \\
\hline
\cite{Cer64} & 1964  & 14.4 & incomplete  & 3,9      & -21.7 $\pm$ 1   \\
\cite{Voi65} & 1965  & 13.9 & incomplete  & 4        & -23.6 $\pm$ 1.8 \\
\cite{Slo68} & 1968  & 13.9 & incomplete  & 2        & -16.7 $\pm$ 2.8 \\
\cite{Bar67} & 1967  & 14.0 & incomplete  & 2        & -14.3 $\pm$ 3   \\
\cite{Shi68} & 1968  & 14.1 & incomplete  & 2        & -18.0 $\pm$ 1.5 \\
\cite{Bon68} & 1968  & 8-28 & incomplete  & 2        & -16.8 $\pm$ 1.0 \\
\cite{Pro70} & 1970  & 14.1 & incomplete  & 4,9      & -23.2 $\pm$ 1.8 \\
\cite{Str72} & 1972  & 50   & incomplete  & 2,9      & -21.7 $\pm$ 1.2 \\
\cite{Shi73a}& 1973  & 14.1 & incomplete  & 2,9      & -19.3 $\pm$ 0.8 \\
\cite{Shi73b}& 1973  & 14.1 & incomplete  & 5,6,7,9  & -18.3 $\pm$ 0.2 \\
\cite{Hai77} & 1977  & 14.0 & incomplete  & 6,7,9    & -23.2 $\pm$ 3.6 \\
\cite{Shi86} & 1986  & 49.6 & incomplete  & 6,7,9    & -19.6 $\pm$ 1.2 \\
\cite{Koo86} & 1986  & 62.8 & incomplete  & 6,7,9    & -18.8 $\pm$ 1.0 \\
\cite{Gra69} & 1969  & 14.1 & complete    & 3,9,11   & -16.2 $\pm$ 2.2 \\
\cite{Zei70} & 1970  & 18.8 & complete    & 1,9,11   & -16.4 $\pm$ 2.8 \\
\cite{Zei73} & 1973  & 18.4 & complete    & 6,7,9,11 & -17.1 $\pm$ 0.8 \\
\cite{Sau72} & 1972  & 14.3 & complete    & 2,9,11   & -25   $\pm$ 3   \\
\cite{Zei74} & 1974  & 18.4 & complete    & 6,7,9,11 & -16.3 $\pm$ 1.0 \\
\cite{Kec75} & 1975  & 14.1 & complete    & 6,7,9,11 & -15.7 $\pm$ 2   \\
\cite{Bre74} & 1974  & 14.2 & complete    & 6,7,9,11 & -16.0 $\pm$ 1.2 \\
\cite{One78} & 1978  & 120  & complete    & 1,9,11   & -17.5 $\pm$ 4   \\
\cite{Wit79a}& 1979  & 17-21 & complete   & 6,7,10,11& -16.9 $\pm$ 0.6 \\
\cite{Wit79b}& 1979  & 21-27 & complete   & 6,7,10,11& -17.0 $\pm$ 0.6 \\
\cite{Gon99} & 1999  & 13.0 & complete    & 6,8,10,12& -18.7 $\pm$ 0.6 \\
\cite{Huh00} & 2000  & 25.2 & complete    & 6,8,10,12& -16.3 $\pm$ 0.4 \\
\hline
\end{tabular}\\[2pt]
\end{table}

\begin{table}[htb]
\caption{This table gives a short list of the main features of the theory used in the analysis
     of {\em nd} breakup data in studies designed to obtain a value for $a_{nn}$ from the
     {\em nn} FSI cross-section enhancement.}
\label{tab:details}
\newcommand{\m}{\hphantom{$-$}}
\newcommand{\cc}[1]{\multicolumn{1}{c}{#1}}
\renewcommand{\tabcolsep}{2pc} 
\renewcommand{\arraystretch}{1.2} 
\begin{tabular}{|rl|}
\hline
            &                    \\
Detail & ~~~~~~~~~~~~~~Description   \\
\hline
  1 & Watson-Migdal and effective range theory \\
  2 & Impulse Approximation, Pole Approximation \\
  3 & Born Approximation \\
  4 & Truncated graph method \\
  5 & Hybrid Final-State-Interaction theory \\
  6 & Faddeev theory \\
  7 & Separable nucleon-nucleon potential \\
  8 & Realistic nucleon-nucleon potential \\
  9 & $\ell = 0$ only NN partial waves \\
 10 & $\ell \geq 0$  NN partial waves \\
 11 & fit shape of {\em nn} FSI enhancement only \\
 12 & fit shape and absolute cross section \\

\hline
\end{tabular}\\[2pt]
\end{table}

While the lists in the tables below are
rather extensive, they are likely not all inclusive.  The omission
of an experiment of the type being reviewed here 
is not to be interpreted as the result of a data evaluation 
exercise but rather as
an oversight on the part of the author. The author apologizes for any
such omissions.  
For each reaction type one or two experiments
that reported relatively small error bars for $a_{nn}$ are discussed
for the purpose of presenting an overview of the experimental techniques
employed and the theory applied in the analysis to 
determine $a_{nn}$ from the measured cross sections. 

\subsection{{\em nd} breakup}

Details of the {\em nd} breakup experiments for which the results are 
plotted in the left and middle sections of Fig.~\ref{fig:ann_summary}
are given in Tables~\ref{tab:ndbu} and \ref{tab:details}.
Typically in the KI experiments 
the momentum vector of the emitted proton, $\vec{p_p}$, is measured.  There are in the
general sense two types of KC experiments.  In the most
common type the deuterium target is an active deuterated scintillator (DS).  
This type will be referred to as KC1.  In 
these experiments the momentum vectors of both emitted neutrons and the energy
of the emitted proton,
$\vec{p_{n1}}$, $\vec{p_{n2}}$, and $E_p$, respectively, are measured. 
Sizeable experimental effects must be taken into account when 
comparing theory to data taken in KC1 experiments.  The most important effects are: neutron
attenuation in the DS, neutron multiple scattering in the DS,
the energy dependence of the efficiency of the two neutron detectors and angle
and energy averaging over the experimental acceptance. Some early high-precision
KC1 experiments were done by B. Zeitnitz {\em et al.} \cite{Zei70,Zei73,Zei74}.  
The experimental
setup is slightly different
in the other type of KC experiment, which will be referred to as KC2.  
In KC2 {\em nd} breakup experiments the deuterium target is a thin foil 
so that the momentum
of the proton, $\vec{p_{p}}$, can be measured.  In these experiments both
$\vec{p_{p}}$ and $\vec{p_{n1}}$ are measured for each detected breakup event.
As with KC1 experiments, the comparision of data taken in KC2
experiments
to theory requires a number of effects to be taken into account:
angle and energy averaging over the experimental acceptance, 
the energy and position dependence of the efficiency of the neutron detector,
and the detection correlation of the two emitted neutrons.  An example of an
early
high-precision KC2 experiment was reported by von Witsch {\em et al.} 
\cite{Wit79a,Wit79b}. 

The results for $a_{nn}$ from the two most 
recent KC experiments of Gonz\'{a}lez Trotter{\em et al.} \cite{Gon99}, which
was the KC1 type, and of 
Huhn {\em et al.} \cite{Huh00}, which was the KC2 type, disagree by more
than three standard deviations of the reported experimental uncertainties. 
This situation is a bit puzzling given that the data taken in both
experiments were analyzed with the same theory.   

As shown in Fig.~\ref{fig:ann_summary} there has been a large number of kinematically
incomplete {\em nd} breakup experiments over the last 35 years with the aim of determining
$a_{nn}$.  While each experiment has unique features, there are some common characteristics.
The main common attributes of these experiments are: (1) the deuterium target is a deuterated
polyethelene foil that is thin enough to allow the emitted protons to pass through with 
sufficient energy for detection, (2) a charged particle detector system is positioned to
detect the protons from the breakup reaction that are emitted around $0^\circ$, and (3) the 
value of $a_{nn}$ is obtained by fitting the enhancement in the proton-energy ($E_p$) spectrum
due to the {\em nn} FSI. The charged-particle detection system is usually either a magnetic
spectrometer or a counter telescope with gas counters to eliminate the direct neutrons
from the event trigger. The main technical challenge in these measurements is the reduction
of protons from background sources, particularly in the flat part of the $E_p$ spectrum.
Even with more than ten kinematically-incomplete {\em nd} breakup experiments contributing 
to the effort, the result reported by 
Shirato {\em et al.} \cite{Shi73b} dominates the computed average for this
type of 
measurement due to the small uncertainty in their value relative to that obtained 
in the experiments. A gas counter telescope was used in the experiment of 
Shirato {\em et al.} \cite{Shi73b}.  The main experimental issue in their
measurement
was the determination of the breakup protons from the lower energy part of their 
incident neutron beam.  These contaminate protons affected the flat part of the $E_p$
spectrum and had little influence on the {\em nn} FSI enhancement, which is at the
extreme high end of the spectrum. In their analysis the $E_p$ spectrum was fit with a 
two-term function by searching on three free parameters, the normalization constant for the 
term that described the flat part of the spectrum, the normalization constant for
the term that described the {\em nn} FSI enhancement region of the spectrum, and $a_{nn}$.
  
The reanalysis of the $E_p$ spectrum from some of the more recent kinematically
incomplete experiments by Tornow {\em et al.} \cite{Tor96} gives very puzzling
results.  In all cases the magnitude of $a_{nn}$ shifted to a substantially smaller
value in the reanalysis with modern theory.  On average the shift was about
3 fm out of 19 fm. The shift seems mostly associated with the discrepancy between 
the cross-section data in the flat part of the proton energy spectrum and the modern
calculations.  Because this part of the energy distribution is insensitive to the value 
of $a_{nn}$, the data were normalized to the calculations in this region. At incident
neutron energies near 14 MeV the data were typically larger than the calculated
cross sections.  The consequence of 
this normalization is that the values of $a_{nn}$ determined in the analysis of 
Tornow {\em et al.} \cite{Tor96} are on average lower than the original values.

\subsection{$\pi^-d$ capture}

The situation for determining $a_{nn}$ from the $\pi^-d$ capture reaction is summarized
in Table~\ref{tab:pi_d}. In these experiments a degraded pion beam is stopped in a
liquid deuterium target.  As in the case of the {\em nd} breakup experiments, the
$\pi^-d$ capture measurements also are divided according to the number of
kinematic parameters measured. 

\begin{table}[htb]
\caption{Survey of $\pi^-d$ capture cross-section measurements that were 
   used to determine $a_{nn}$.  The survey covers the period between 1964
   and the present.}
\label{tab:pi_d}
\newcommand{\m}{\hphantom{$-$}}
\newcommand{\cc}[1]{\multicolumn{1}{c}{#1}}
\renewcommand{\tabcolsep}{2pc} 
\renewcommand{\arraystretch}{1.2} 
\begin{tabular}{|cllcl|}
\hline
          &       & \cc{Measured}   & \cc{Analysis}  & $a_{nn} \pm \Delta a_{ann}$ \\
    ref.  & year  & \cc{Kinematics} & \cc{ref.}      & ~~~~(fm)                   \\
\hline
\cite{Had65} & 1965  & complete   & \cite{Ban64} & -16.4 $\pm$ 1.9 \\
\cite{Sal75} & 1975  & complete   & \cite{Ban64} & -16.7 $\pm$ 1.3 \\
\cite{Gab79} & 1979  & incomplete & \cite{Ter77} & -18.5 $\pm$ 0.4 \\
\cite{Sch87} & 1987  & complete   & \cite{Ter77} & -18.7 $\pm$ 0.6 \\
\cite{How98} & 1998  & complete   & \cite{Gib75} & -18.5 $\pm$ 0.5 \\

\hline
\end{tabular}\\[2pt]

\end{table}

\noindent
In KI measurements of the $\pi^- + d \rightarrow 2n + \gamma$
reaction only the energy of the $\gamma$-ray ($E_\gamma$) is measured. The 
measured $E_\gamma$ spectrum is
fit with theory to determine a value of $a_{nn}$.  In the KC meaurements, the 
momenta of the $\gamma$-rays and one of the neutrons are measured for
each detected capture event.  The time-of-flight (TOF) spectrum of the detected neutron
is fit to determine $a_{nn}$. The main experimental effects that must be taken into 
account when fitting the neutron TOF spectrum are the energy dependence of the neutron
detection efficiency, the energy dependence of the attenuation and scattering of 
neutrons in the liquid
deutrium target and the surrounding materials. The results from the two most recent
KC experiments are in agreement within the reported uncertainties, which include a
$\pm$0.3~fm due to theoretical uncertainties.

\section{SUMMARY and RECOMMENDATIONS}

The most popularly accepted value for $a_{nn}$ comes from the $\pi^-d$ capture measurements.
The general belief is that these results are more reliable than those obtained from
{\em nd} breakup or from other reactions in which there are three or more hadrons 
in the exit channel.  This supposition is supported by the observation that the $a_{nn}$ values
obtained in the most recent high-precision $\pi^-d$ experiments are in agreement while
recent {\em nd} breakup experiments and analyses give discrepant results.  Therefore,
the recommended value for $a_{nn}$ obtained using indirect methods is -18.6 $\pm$0.3 
(experimental) $\pm$0.3 (theory) fm. 

An important next step is to conduct {\em nd} breakup experiments for the purpose of
resolving the discrepancies between the results from recent KC {\em nd} breakup experiments
and for determining the cause of the shift in the value of $a_{nn}$ in the
reanalysis of the KI {\em nd} breakup cross-section data.  
While it is unclear whether investigating these problems will lead to a better determination
of $a_{nn}$, this work will almost certainly strenghten our understanding of three-nucleon
reaction dynamics in the kinematic region around the {\em nn} FSI. 

The next sufficient step in this problem would be a direct measurement of $a_{nn}$.  The
high thermal neutron flux at the YAGUAR pulsed reactor in Russia 
opens opportunties for such measurements.  The DIANNA collaboration is planning the
first direct measurement of $a_{nn}$ that use neutrons from a reactor.  Some details of 
their proposed
experiment are given in these proceedings.

\end{document}